\documentstyle[12pt]{article}

\newcommand{\be}{\begin{equation}}
\newcommand{\ee}{\end{equation}}
\newcommand{\bea}{\begin{eqnarray}}
\newcommand{\eea}{\end{eqnarray}}
\newcommand{\vs}[1]{\vspace{#1 mm}}

\renewcommand{\b}{\beta}

\newcommand{\e}{\epsilon}

\def\bbox{{\,\lower0.9pt\vbox{\hrule \hbox{\vrule height 0.2 cm
\hskip 0.2 cm \vrule height 0.2 cm}\hrule}\,}}
\newcommand{\dsl}{\pa \kern-0.5em /}

\newcommand{\pa}{\partial}

\newcommand{\nn}{\nonumber\\}

\font\mybb=msbm10 at 10pt
\def\bb#1{\hbox{\mybb#1}}

\def\bR {\bb{R}}
\def\bE {\bb{E}}

\def\b{\beta}

\def\e{\epsilon}           
\def\f{\phi}               

\def\o{\omega}  

\def\D{\Delta}

\def\G{\Gamma}

\def\L{\Lambda}

\def\cc{{\cal C}}

\def\cn{{\cal N}}

\def\half{{1 \over 2}}

\begin{document}

\topmargin 0pt
\oddsidemargin 5mm

\renewcommand{\thefootnote}{\fnsymbol{footnote}}
\begin{titlepage}

\setcounter{page}{0}
\begin{flushright}
DAMTP-1999-111\\
SPIN-1999/20\\
hep-th/9909070
\end{flushright}

\vs{5}
\begin{center}
{\Large \bf Gravitational Stability and Renormalization-Group Flow}
\vs{12}

{\large {\bf
Kostas Skenderis}\footnote{e-mail address:K.Skenderis@phys.uu.nl}
} \\
\vs{3}
{\em Spinoza Institute, University of Utrecht \\
Leuvenlaan 4, 3584 CE Utrecht \\
The Netherlands} \\
\vs{2}
and \\
\vs{2}
{\large {\bf
Paul K. Townsend}\footnote{e-mail address: pkt10@damtp.cam.ac.uk}
} \\
\vs{3}
{\em DAMTP, University of Cambridge \\
Silver Street, Cambridge CB3 9EW, UK}\\ 
\end{center}
\vs{7}
\centerline{{\bf{Abstract}}}

First-order `Bogomol'nyi' equations are found for dilaton domain walls
of D-dimensional gravity with the general dilaton potential admitting a stable
anti-de Sitter vacuum. Implications for renormalization group flow in the
holographically dual field theory are discussed. 

\end{titlepage}
\newpage
\renewcommand{\thefootnote}{\arabic{footnote}}
\setcounter{footnote}{0} 

\section{Introduction}

The strong t' Hooft-coupling limit of certain non-conformal supersymmetric 
quantum field theories associated with coincident non-conformal
branes has a description in terms of supergravity theory \cite{It}. 
This description involves gauged supergravities admitting 
domain-wall vacua \cite{BST}.  
The Minkowski vacuum of the gauge theory at a given scale is a
`horosphere' of a supergravity dilaton domain wall, i.e. a hypersurface in the
`holographic frame' anti-de Sitter (adS) metric on which the dilaton is
constant \cite{BST}. The position of the horosphere and 
the value of the dilaton is directly related to the energy scale of the 
gauge theory.
The domain wall solution itself therefore corresponds in
the gauge theory to renormalization-group (RG) flow from one 
scale to another. The
cases considered in \cite{BST}, and similar lower-dimensional cases
\cite{bergs}, are all ones for which the dilaton potential is a simple
exponential. In such cases there is no maximally-supersymmetric adS vacuum but
there is a 1/2 supersymmetric linear-dilaton vacuum which can be interpreted as
a domain wall. Another
type of domain wall, interpolating between adS vacua with different radii of
curvature, has been extensively studied in the context of D=4 supergravity
\cite{Cvetic,DWreview}, and similar solutions have recently been found for D=5
supergravity theories \cite{GPPZ1,DZ,Freedman,Beh}. 
These domain walls correspond to RG 
flow from one superconformal field theory to another. 
Other examples of RG flows of 
$d=4, \cn=4$ SYM theory that have a description in terms of 
D=5 supergravity can be found in \cite{KeSf,Gu,GPPZ2,FGPW,SB,GPPZ3}.
More recently, the
RG flow associated with domain walls has been used in the context of `Brane
World' scenarios to explain the origin of mass hierarchies and as a possible
explanation for the smallness of the cosmological constant \cite{RS,V,VE}.

Given these new applications of domain wall spacetimes, it would be helpful to
have a model-independent analysis of the possibilities in which basic
physical requirements are the only input. Since matter fields other than
scalars play no role in domain wall solutions, the general framework is
gravity coupled to a scalar field theory in $D$ spacetime dimensions. The
scalar fields will take values in some target space ${\cal M}$ and the model is
characterized by the metric on ${\cal M}$, which determines the scalar
kinetic terms, and a function $V$ on ${\cal M}$, which determines the scalar
potential. The target space metric must be positive definite for vacuum
stability. Intuition from non-gravitational field theory might lead one to
suppose that vacuum stability also requires that $V$ be positive but in gauged
supergravity theories $V$ is typically unbounded from below, and the
supersymmetric adS vacua are either maxima or saddle points of $V$
\cite{Warner}. The perturbative stability of these adS vacua is guaranteed by 
the
fact that the eigenvalues of the scalar mass matrix satisfy
the Breitenlohner-Freedman bound \cite{BF} or its D-dimensional generalization
\cite{MT}. Non-perturbative stability has also been established in many cases 
by
an extension of the spinorial proof of the positive energy theorem
\cite{Witten} to asymptotically adS spacetimes \cite{GHW,MTvN}. This method was
used in \cite{Boucher,PKT} to determine the restrictions on $V$ that arise from
the requirement that there exist a stable adS vacuum, whether supersymmetric or
not. The results imply the perturbative stability bounds of \cite{BF,MT} but go
well beyond them by providing information about the potential {\sl away from 
its critical points}. 

This information is particularly useful if one supposes that
there is only a single scalar field $\phi$, which we shall call the
`dilaton'. In this case ${\cal M}=\bR$ so the target space metric is
diffeomorphic to a constant and $V$ becomes a function of a single
real variable. The general model discussed above reduces to one with  
Lagrangian density
\be\label{lag}
{\cal L} = \sqrt{-\det g}\left[{1\over2} R - {1\over2}(\partial\phi)^2
-V(\phi)\right]
\ee
where $g$ is the D-dimensional spacetime metric with `mostly plus' signature. 
The result of \cite{PKT} is that vacuum stability requires $V$ to take the form
\be\label{potential}
V= 2(D-2)\left[(D-2)(w')^2 - (D-1)w^2\right]
\ee
where $w(\phi)$ is any function admitting at least one critical point
and the prime indicates differentiation with respect to $\phi$. We
shall call $w(\phi)$ the `superpotential'. The restriction 
to a single scalar
field might appear severe but there are many supergravity theories of interest
for which there is only one scalar or for which it is natural to consider the
truncation to a single scalar. For example, in all effective supergravity
theories associated to string theory there is a natural truncation in which 
only
the dilaton survives; hence our choice of terminology. There are also some 
cases
in which the potential depends only on the dilaton even though this is not the
only scalar field. In many such cases the potential is given by the above
formula with $\log w \propto \phi$ even though this superpotential has no
critical point. This suggests that the formula (\ref{potential}) is valid under
conditions less restrictive than those used in its derivation. 
Notice also that a potential 
of the form (\ref{potential}) 
for the multi-scalar case still 
guarantees gravitational stability \cite{PKT} although the converse is not 
necessarily true, i.e. in the multi-scalar case there may be more general 
potentials than (\ref{potential}) compatible with gravitational stability.
In particular, the potential of a subset of the scalars of the 
D=5 supergravity used in recent studies \cite{Freedman,Beh,FGPW}
is of the form (\ref{potential}). 
The potential (\ref{potential}) 
has a form that is typical in supergravity theories, hence the choice
of terminology `superpotential' for $w$, even though supersymmetry is {\sl not}
an ingredient in its derivation. 

In this paper we will investigate general properties of domain wall solutions
in the theory with Lagrangian (\ref{lag}) with $V$ given by (\ref{potential}). 
Our interest in domain wall spacetimes stems from their connection to 
the RG flow of the dual field theories. 
Such models are characterized by their superpotential $w$. Let us first note
that
\be
V' = 4(D-2)\left[(D-2)w'{}' -(D-1)w\right]\, w'
\ee
so that $V$ has critical points at critical points of $w$, and at points for
which
$w'' = {D-1\over D-2} w$.
In the context of supergravity theories the critical points of $w$ yield
stable adS vacua. The other critical points of $V$ yield 
non-supersymmetric (but usually adS) vacua which may or may not be
stable. Recall that the positivity of the energy, and hence stability,
is established subject to prescribed boundary conditions at infinity,
so the fact that $V$ as given in (\ref{potential}) was derived by
requiring the existence of a stable adS vacuum does not imply that all
of its adS vacua are stable; each such vacuum requires its own
boundary conditions. 

In addition to adS vacua there will usually be domain wall solutions. 
These solutions are possible, and may even be supersymmetric,
regardless of whether $V$ has critical points. If $V$ does have 
critical points then some of these domain wall solutions will 
interpolate between the corresponding adS vacuum and some other 
solution, possibly another adS vacuum. It is convenient to distinguish
between two types of domain wall, the `BPS' ones and the 'non-BPS'
ones. In the supergravity context the BPS walls are
supersymmetric, and they interpolate between supersymmetric
vacua. Domain walls that interpolate between a supersymmetric vacuum
and a non-supersymmetric one, or between two non-supersymmetric vacua,
are necessarily non-BPS. Our focus will be on BPS domain walls, but we
shall first consider the general case.

\section{Domain walls and the c-function}

We begin by making the domain-wall ansatz
\be \label{domain}
ds^2= e^{2 A(r)} ds^2\left(\bE^{(1,D-2)}\right) + dr^2
\ee
with dilaton field $\phi(r)$. 
Let us introduce a new radial coordinate 
$U=e^{A}$. The domain-wall spacetime then takes the form
\be \label{Uform}
ds^2 = U^2 ds^2\left(\bE^{(1,D-2)}\right) + 
{1 \over (\pa_r A)^2} {dU^2 \over U^2}\, .
\ee
At critical points of $V$ the dilaton is constant, as is $(\partial_r
A)$, and the geometry becomes anti-de Sitter with a cosmological
constant $\Lambda$ equal to the value of $V$ at the critical point; 
$\L=-\half (D-1) (D-2) (\pa_r A)^2$. In the dual field theory this
corresponds to a conformal fixed point of the RG flow.
The variable $U$ is identified with the renormalization-group scale; 
$U=\infty$ corresponds to long distances in the bulk, so UV in 
the dual field theory, and $U=0$ to short distances in the bulk, so 
IR in the dual field theory.
The RG flow of the coupling constant(s)
of the field theory is encoded in the $U$ dependence of the scalar field(s).
At a fixed point the scalar field is constant, and 
the corresponding $\b$-function vanishes.  

The Einstein-dilaton equations for the metric (\ref{domain}) reduce to
\bea
&&(D-2)(D-1)(\pa_r A)^2 - (\pa_r \f)^2 + 2 V(\f)=0 \label{secorder1} \\
&&2 (D-2) \pa_r^2 A + (D-1) (D-2) (\pa_r A)^2
+ (\pa_r \f)^2 + 2 V(\f) =0 \label{secorder2} \\
&&\pa_r^2 \f + (D-1) \pa_r A \pa_r \f - V'(\f) = 0 \label{secorder3}
\eea
where the prime again indicates differentiation with respect to $\phi$. 
Not all three equations are independent, however. For instance, 
one can obtain (\ref{secorder2}), upon differentiation of (\ref{secorder1})
and using (\ref{secorder1}) and (\ref{secorder3}). 
These equations imply 
\be \label{sec4}
\pa_r^2 A = - {1 \over D-2} (\pa_r \f)^2.
\ee

In \cite{GPPZ1,Freedman} the function 
\be
\cc(U) = \cc_0/\left[\partial_r A(r)\right]^{D-2}
\ee
was proposed as a c-function, where $\cc_0$ is a constant related
to the universal coefficient appearing in the ``holographic''
Weyl anomaly \cite{HS} (for odd $D$).
By definition, a c-function is a positive
function of the coupling constant(s) that is non-increasing along the
RG flow from the UV to the IR. We can easily 
establish monotonicity:
\be
U {\partial \over \partial U} \cc = 
- (D-2) \cc {1 \over (\pa_r A)^2} \pa_r^2 A=
\cc \left({ \pa_r \f \over \pa_r A} \right)^2 \geq 0,
\ee
where we have used (\ref{sec4}). Thus,
as we move from the UV at $U=\infty$ towards 
the IR at $U=0$, $\cc$ is non-increasing.  
More generally, it was shown 
in \cite{Freedman} that the function $\cc(U)$ is monotonic
as a consequence of a `weaker' energy condition on the bulk matter.

\section{BPS domain walls}

The equations (\ref{secorder2})-(\ref{secorder3}) 
are the Euler-Lagrange equations for the 
functional
\be\label{energy}
E[A,\phi] = {1\over 2}\int_{-\infty}^\infty\! dr\, 
e^{(D-1)A}\left[(\partial_r\phi)^2 -
(D-1)(D-2)(\partial_r A)^2 + 2V\right]\, .
\ee
The integrand is minus the effective Lagrangian obtained by substitution of
the domain wall ansatz into the Lagrangian (\ref{lag}), and the functional
$E$ is simply related to expressions for the energy obtained by other
means\footnote{For example, use of the field equations allows $E$ to
be expressed as a surface integral of the second fundamental form,
which was shown in \cite{HH} to be proportional to the energy.}.

The functional (\ref{energy}) can be rewritten, 
{\` a} la Bogomol'nyi, as 
\bea
E &=& {1\over2}\int_{-\infty}^\infty\! dr \, 
e^{(D-1)A}\left\{ [\partial_r\phi \mp
2(D-2)w']^2 -(D-1)(D-2)[\partial_r A \pm 2w]^2\right\} \nonumber\\
&& \qquad \pm\, 2(D-2)[e^{(D-1)A}w]_{-\infty}^\infty\, .
\eea
It follows that $E$ is extremised by solutions of the
following pair of first-order equations:
\bea \label{cond}
\partial_r A &=& \mp 2 w(\f) \nn
\partial_r \phi  &=& \pm 2(D-2) w'(\phi)\, .
\eea
It is straightforward to verify that
solutions of these equations indeed solve the second-order equations
(\ref{secorder1})-(\ref{secorder3}).  
We shall call solutions of these equations
`BPS domain walls'. 

Another way to arrive at the first-order equations (\ref{cond}) is to note
that the energy bound established in \cite{PKT} is
saturated by field configurations for which the equations
\be\label{susyone}
(D_m + w\Gamma_m)\epsilon =0\, ,\qquad 
\left[\Gamma^m\partial_m\phi - 2(D-2)w'\right]\epsilon =0\, ,
\ee
admit solutions for a non-zero spinor $\epsilon$, which we shall 
call a Killing
spinor. Such field configurations automatically solve the second-order
Einstein-dilaton equations. 
Substitution of the domain wall ansatz into the equations (\ref{susyone})
leads immediately to the equations (\ref{cond}). The Killing spinor is
$\e = e^{A/2} \e_0$ with $\e_0$ a constant spinor satisfying 
$\G_r \e_0 = \pm \e_0$. In the context of
supergravity, the domain walls admitting Killing spinors are
supersymmetric.

\section{Examples}

We now consider two examples involving the $N=1, D=7$
supergravity \cite{TvN} and the $N=2, D=6$ $F(4)$ supergravity
\cite{romans}. The D=7 theory is obtained by compactification of D=11
supergravity on $S^4$ \cite{LP}, and it is associated with
the near-horizon limit of the $M5$ brane with an orbifold 
projection on the transverse sphere \cite{berkooz,FKPZ1,AOT}. 
The D=6 theory is
obtained by a warped $S^4$-compactification of massive IIA
supergravity \cite{CLP}, and it is associated with 
the near-horizon limit of the 
D4-D8 system \cite{FKPZ2,BO}. In both cases there is a single scalar 
field $\f$, and we may discuss them simultaneously.
The potential is given by (\ref{potential}) with
\be
w(\f) = - {1 \over 2 \sqrt{2} (D-2)}
\left(g e^{{1 \over \sqrt{D-2}} \f} 
+ {m \over \sqrt{D-5}} e^{-{D-3 \over \sqrt{D-2}} \f} \right)
\ee
Here $g$ is the coupling constant of the gauge group and $m$ is a
`topological' mass parameter. The potential has two critical points:
at $w'=0$ and at $w'' = {D-1\over D-2} w$. The two critical points are
\be
e^{\sqrt{D-2}\f}={m \over g} {D-3 \over \sqrt{D-5}}, 
\qquad e^{\sqrt{D-2}\f}={m \over g} \sqrt{D-5},
\ee
Only the first ($w'=0$) critical point is supersymmetric. 

Domain-wall solutions of these supergravity theories, preserving 1/2
supersymmetry, were found in \cite{LPS}. In terms of a new radial
parameter $\rho$, they take the form
\be
ds^2 = e^{{2B \over D-3}} ds^2(\bE^{(1,D-2)}) + e^{2 B} d\rho^2, 
\qquad \phi = {\sqrt{D-2}\over (D-4)}\, \log \rho\, ,
\ee
where
\be  
e^{-B}=2\rho (D-4) \sqrt{D-2}\, w'(\f)\, .
\ee
There is an apparent singularity at the critical point $w'=0$ but
this is only a coordinate singularity, as one can verify by choosing
$U=e^{{B \over D-3}}$ as a new radial variable. The metric then becomes
\be
ds^2 = U^2 ds^2(\bE^{(1,D-2)}) + \left({D-3 \over \pa_\rho e^{-B}}\right)^2 
{dU^2 \over U^2}\, ,
\ee
which is non-singular when $w'=0$. The relation  
\be
\left[w'' - {D-3 \over D-2} w\right]_{w'=0} =0
\ee
is required in order for the domain wall solution to become the 
supersymmetric adS
solution as the critical point is approached. This relation
turns out to be satisfied.
In the new radial coordinate the critical point is at $U=\infty$, 
so it corresponds to a UV fixed point of the dual field theory.

We conclude with an example of a superpotential admitting a
BPS domain wall but which is not the superpotential of
any known supergravity theory (at least not for general $D$).
A class of solutions of  equations (\ref{cond}) is obtained by 
first considering these equations for complex $\f$, $w(\f)$, $A(r)$, 
and then imposing reality conditions\footnote{
This method has also been recently used in \cite{BSf} in 
order to obtain supersymmetric domain wall solutions 
of D=5 supergravity.}. As an example we consider 
the case the superpotential $w(\f)$ is equal to the Weierstrass 
elliptic function, $w(\f)=\wp(\f;g_2,g_3)$. 
The dilaton $\f$ is then the uniformizing variable of the 
elliptic curve associated to the Weierstrass function.
Let us recall 
some standard facts about the Weiersrass function, $\wp(\f;g_2,g_3)$.
It satisfies the differential equation,
\be \label{wei}
\wp'^2=4 \wp^3 - g_2 \wp - g_3.
\ee
It follows that the superpotential has three critical points. 
The value of the superpotential at the critical points 
is given by the three roots, $e_1, e_2, e_3$, of the cubic
polynomial in the right hand side of (\ref{wei}). 
The critical
points occur at $\f=\o_1, \f=\o_1+\o_2, \f=\o_2$, where 
$\o_1$ and $\o_2$ are the half periods of $\wp$.

One can easily integrate equations (\ref{cond}). The result is 
\bea
r-r_0&=&\pm {1 \over 8 (D-2)} \left[
{1 \over (e_1 - e_2)(e_1-e_3)} \log (\wp - e_1) \right. \nn
&+&\left.  {1 \over (e_2 - e_1)(e_2-e_3)} \log (\wp - e_2) +
{1 \over (e_3 - e_1)(e_3-e_2)} \log (\wp - e_3) \right] \nn
A(r(\f))-A_0&=&-{1 \over 4 (D-2)}
\left[{e_1 \over (e_1 - e_2)(e_1-e_3)} \log (\wp - e_1) \right.\nn
&+&\left. 
{e_2 \over (e_2 - e_1)(e_2-e_3)} \log (\wp - e_2) +
{e_3 \over (e_3 - e_1)(e_3-e_2)} \log (\wp - e_3) \right]
\eea
where $r_0$ and $A_0$ are integration constants, which we set to 
zero so that the critical points occur for $r = \pm \infty$, 
and $U=0$ or $U=\infty$.

When $g_2, g_3$ are real one can impose reality conditions
on the solution. There are two cases to consider. When 
the discriminant $\D=g_2^3-27 g_3^2$ is positive, one may 
choose primitive periods such that $\o_1$ is real and 
$\o_2$ is imaginary. In this case all three roots
$e_i$ ($e_1>e_2>e_3, e_1>0, e_3<0$) are real. 
The (real) superpotential has one critical point at $\f=\o_1$
(the other two critical points occur for complex
values of the dilaton).
When $\D<0$, one may choose $\o_1$, $\o_2$ to be complex
conjugate of each other. The roots $e_1$ and 
$e_3$ are complex conjugates and $e_2$ is real.
The (real) superpotential has one critical point at $\f=\o_1+\o_2$. 
When two of the roots coincide, or what is the same, one of the 
periods becomes infinite, the Weierstrass function reduces
to an elementary function.

\vskip 1cm
\noindent
{\bf Acknowledgments}: We thank Harm Jan Boonstra for 
collaboration at early stages of this work. 
KS is supported by the Netherlands Organization for Scientific
Research (NWO).

\newcommand{\NP}[1]{Nucl.\ Phys.\ {\bf #1}}
\newcommand{\AP}[1]{Ann.\ Phys.\ {\bf #1}}
\newcommand{\PL}[1]{Phys.\ Lett.\ {\bf #1}}
\newcommand{\CQG}[1]{Class. Quant. Gravity {\bf #1}}
\newcommand{\CMP}[1]{Comm.\ Math.\ Phys.\ {\bf #1}}
\newcommand{\PR}[1]{Phys.\ Rev.\ {\bf #1}}
\newcommand{\PRL}[1]{Phys.\ Rev.\ Lett.\ {\bf #1}}
\newcommand{\PRE}[1]{Phys.\ Rep.\ {\bf #1}}
\newcommand{\PTP}[1]{Prog.\ Theor.\ Phys.\ {\bf #1}}
\newcommand{\PTPS}[1]{Prog.\ Theor.\ Phys.\ Suppl.\ {\bf #1}}
\newcommand{\MPL}[1]{Mod.\ Phys.\ Lett.\ {\bf #1}}
\newcommand{\IJMP}[1]{Int.\ Jour.\ Mod.\ Phys.\ {\bf #1}}
\newcommand{\JP}[1]{Jour.\ Phys.\ {\bf #1}}


\begin{thebibliography}{99}

\bibitem{It}
N. Itzhaki, J.M. Maldacena, J. Sonnenschein and S. Yankielowicz, {\sl
Supergravity and the large N limit of theories with sixteen
supercharges}, Phys. Rev. {\bf D58} (1998) 046004.

\bibitem{BST}
H.J. Boonstra, K. Skenderis and P.K. Townsend, {\sl The
domain-wall/QFT correspondence}, JHEP {\bf 9901} (1999) 003, hep-th/9807137;
K. Skenderis, {\sl Field theory limit of branes and gauged supergravities},
hep-th/9903003. 

\bibitem{bergs}
K. Behrndt, E. Bergshoeff, R. Halbersma and J.P. van de Schaar, {\sl
On domain wall/QFT dualities in various dimensions}, hep-th/9907006.

\bibitem{Cvetic}
M. Cveti{\v c}, S. Griffies and S-J Rey, {\sl Static domain walls in N=1
supergravity}, Nucl. Phys. {\bf B381} (1997) 301.

\bibitem{DWreview}
M. Cveti{\v c} and H.H. Soleng, {\sl Supergravity domain walls},
Phys. Rept. {\bf 282} (1997) 159.


\bibitem{GPPZ1} L. Girardello, M. Petrini, M. Porrati and A. Zaffaroni,
{\sl Novel Local CFT and Exact Results on Perturbations of N=4 Super Yang
Mills from AdS Dynamics}, JHEP {\bf 9812} (1998) 022, hep-th/9810126.

\bibitem{DZ} J. Distler and F. Zamora, {\sl Non-supersymmetric Conformal Field
Theories from Stable Anti-de Sitter Spaces}, 
Adv. Theor. Math. Phys. {\bf 2} (1998) 1405-1439, hep-th/9810206. 

\bibitem{Freedman}
D.Z. Freedman, S.S. Gubser, K. Pilch and N.P. Warner, {\sl
Renormalization group flows from holography--supersymmetry and a
c-theorem}, hep-th/9904017.

\bibitem{Beh}
K. Berndt, {\sl Domain walls of D=5 supergravity and fixpoints of N=1
super Yang-Mills}, hep-th/9907070. 

\bibitem{KeSf} A. Kehagias and K. Sfetsos, 
{\sl On the runnings in gauge theories
from type IIB supergravity}, Phys. Lett. {\bf B454} (1999)
270-276, hep-th/9902125.

\bibitem{Gu} S. Gubser, {\sl Dilaton driven confinement}, hep-th/9902155.

\bibitem{GPPZ2}  L. Girardello, M. Petrini, M. Porrati and A. Zaffaroni,
{\sl Confinement and Condensates Without Fine Tuning in Supergravity Duals 
of Gauge Theories}, JHEP {\bf 9905} (1999) 026, hep-th/9903026.


\bibitem{FGPW}
D.Z. Freedman, S.S. Gubser, K. Pilch and N.P. Warner,
{\sl Continuous distributions of D3-branes and gauged supergravity},
hep-th/9906194.

\bibitem{SB} A. Brandhuber and K. Sfetsos, {\sl Wilson loops from 
multicentre and rotating branes, mass gaps and phase structure in 
gauge theories}, hep-th/9906201.

\bibitem{GPPZ3}  L. Girardello, M. Petrini, M. Porrati and A. Zaffaroni,
{\sl The Supergravity Dual of N=1 Super Yang-Mills Theory}, hep-th/9909047.

\bibitem{RS}
L. Randall and R. Sundrum, {\sl A large mass hierarchy from a small extra
dimension}, hep-ph/9905221; {\sl An alternative to compactification},
hep-th/9906064.

\bibitem{V}
H. Verlinde, {\sl Holography and Compactification}, hep-th/9906182.

\bibitem{VE} E. Verlinde, {\sl Holography, Compactification and 
the Cosmological Constant}, talk in String '99, 
http://strings99.aei-potsdam.mpg.de. 

\bibitem{Warner}
N.P. Warner, {\sl Some properties of the scalar potential in gauged
supergravity theories},
Nucl. Phys. {\bf B231} (1984) 250.

\bibitem{BF}
P. Breitenlohner and D.Z. Freedman, 
{\sl Positive energy in anti-de Sitter backgrounds and gauged
extended supergravity},
Phys. Lett. {\bf 115B} (1982) 197; 
{\sl Stability in gauged extended supergravity},
Ann. Phys. {\bf 144} (1982) 249.

\bibitem{MT}
L. Mezincescu and P.K. Townsend, {\sl Stability at a local maximum in higher 
dimensional anti-deSitter space, and applications to supergravity}, Ann. Phys.
(N.Y.) {\bf 160} (1985) 406.

\bibitem{Witten}
E. Witten, 
{\sl A simple proof of the positive energy theorem},
Commun. Math. Phys. {\bf 80} (1981) 381.

\bibitem{GHW}
G.W. Gibbons, C.M. Hull and N. Warner, {\sl The stability of 
gauged supergravity},
Nucl. Phys. {\bf B218} (1983) 173.

\bibitem{MTvN}
L. Mezincescu, P.K. Townsend and P. van Nieuwenhuizen, {\sl Stability of 
gauged d=7 supergravity and the definition of masslessness in $(adS)_7$}, Phys.
Lett. {\bf 143B} (1984) 384.

\bibitem{Boucher}
W. Boucher, {\sl Positive energy without supersymmetry}, 
Nucl. Phys. {\bf B242} (1984) 282.

\bibitem{PKT}
P.K. Townsend, {\sl Positive energy and the scalar potential in higher 
dimensional (super)gravity theories}, Phys. Lett. {\bf 148B} (1984) 55.
 
\bibitem{HS} M. Henningson and K. Skenderis, {\sl The Holographic
Weyl anomaly}, JHEP {\bf 9807} (1998) 023, hep-th/9806087;
{\sl Holography and the Weyl anomaly}, hep-th/9812032.

\bibitem{HH}
S.W. Hawking and G.T. Horowitz, {\sl The gravitational hamiltonian,
action, entropy and surface terms}, Class. Quantum Grav. {\bf 13}
(1996) 1487-1498, gr-qc/9501014.

\bibitem{TvN} P.K. Townsend and P. van Nieuwenhuizen,
{\sl Gauged seven-dimensional supergravity}, Phys. Lett. {\bf B125} 
(1983) 41.

\bibitem{romans} L.J. Romans, {\sl The F(4) gauged supergravity
in six dimensions}, Nucl. Phys. {\bf B269} (1986) 691.

\bibitem{LP}
H. L\"u and C.N. Pope, {\sl Exact embedding of N=1 D=7 gauged
supergravity in D=11}, hep-th/9906168.

\bibitem{berkooz}
M. Berkooz, {\sl A Supergravity Dual of a (1,0) Field Theory 
in Six Dimensions}, Phys. Lett. {\bf B437} (1998) 315-317,
hep-th/9802195.

\bibitem{FKPZ1} 
S. Ferrara, A. Kehagias, H. Partouche and A. Zaffaroni,
{\sl Membranes and Fivebranes with Lower Supersymmetry and their AdS 
Supergravity Duals}, Phys. Lett. {\bf B431} (1998) 42-48, hep-th/9803109.

\bibitem{AOT}
C. Ahn, K. Oh and R. Tatar,
{\sl Orbifolds $AdS_7 \times S^4$ and Six Dimensional (0, 1) SCFT},
Phys. Lett. {\bf B442} (1998) 109-116, hep-th/9804093.

\bibitem{CLP}
M. Cveti{\v c}, H. L\"u and C.N. Pope, {\sl Gauged six-dimensional
supergravity from massive Type IIA}, hep-th/9906221.

\bibitem{FKPZ2} 
S. Ferrara, A. Kehagias, H. Partouche and A. Zaffaroni,
{\sl $AdS_6$ Interpretation of 5d Superconformal Field Theories},
Phys. Lett. {\bf B431} (1998) 57-62, hep-th/9804006.

\bibitem{BO} A. Brandhuber and Y. Oz,
{\sl The D4-D8 Brane System and Five Dimensional Fixed Points},
Phys. Lett. {\bf B460} (1999) 307-312, hep-th/9905148.

\bibitem{LPS} H. L\"{u}, C.N. Pope, E. Sezgin and K.S. Stelle,
{\sl Dilatonic p-brane solutions}, Phys. Lett. {\bf B371} (1996) 46-50,
hep-th/9511203.

\bibitem{BSf} I. Bakas and K. Sfetsos, 
{\sl States and Curves of Five-Dimensional Gauged Supergravity},
hep-th/9909041.


\end{thebibliography}
\end{document}